\documentclass[12pt]{article}
\usepackage{epsfig}

\newcommand\email[1]{}
\newcommand\institution[1]{}
\newcommand\address[1]{}
\begin{document}

\begin{flushright}
{HU-EP-07/24}
\end{flushright}
\begin{center}
{\baselineskip=24pt {\Large \bf
Confinement encoded in Landau gauge \\ gluon and ghost propagators
}\\
\vspace{0.5cm}

{\large E.-M.~Ilgenfritz}

\vspace{0.5cm} {\baselineskip=16pt { \it
Humboldt-Universit\"at zu Berlin, Institut f\"ur Physik, \\
	Newtonstrasse~15, 12489 Berlin, Germany} }}
\end{center}

\vspace{0.5cm}
\abstract{An introduction is given into current lattice investigations of the 
non-perturbative gluon and ghost propagators, in the light of the 
Gribov-Zwanziger and Kugo-Ojima scenarios of confinement, in the context of 
results obtained from the non-perturbative Dyson-Schwinger approach in the 
continuum and in connection with the vortex mechanism of confinement.}

\vspace{0.5cm}
\section{Introduction}

In lattice gauge field theory, confinement of quarks is numerically proven 
although the dynamical origin is still under debate~\cite{Alkofer:2006fu}.
An exponential area law holds for Wilson loops in the fundamental representation. 
What about gluon confinement ? Wilson loops with (static) adjoint charges 
do not decay with an area law, but gluons are confined, too. This talk is an 
introduction to alternative confinement 
ideas~\cite{Gribov:1977wm,Zwanziger:1991ac,Kugo:1979gm} 
and presents a report on combined efforts by continuum and lattice theorists to 
understand how they might be realized in Nature.

The infrared behavior of gluon, ghost and quark propagators 
is the focus of a field-theoretic approach~\cite{Alkofer:2000wg} 
to confinement. Green's functions 
carry all information about the structure of a theory. These propagators, 
in distinction to hadron propagators, are gauge-variant. This is the origin 
of difficulties related to the {\it Gribov ambiguity} which shows up at different places. 
The first, pioneering study~\cite{Mandula:1987rh} of the gluon propagator 
in Landau gauge (on lattices as small as $4^3\times8$) dates back to 1987.
Gluon and ghost propagators became topics of stronger interest in the middle 
of the 90-s. A first review about this activity was given in 
\cite{Mandula:1999nj}. 

To establish a relation between confinement and the gluon and ghost propagators 
one mostly concentrates on the infrared momentum range. Is this range,
where the asymptotic behavior sets in, $O(100)$ or $O(10)$ MeV or smaller ? 
In order to probe small momenta, one needs to control the infinite-volume limit.
This makes the problem difficult on the lattice, even with present day
lattice sizes and computers. The following are the signatures of confinement 
from this point of view :
\begin{itemize}
\item The gluon propagator should vanish in the limit 
$q \to 0$~\cite{Smekal:1997is,Smekal:1998vx}. 
The gluon dressing function $Z(q^2)$, defined through
\begin{equation}
D^{ab}_{\mu\nu}(q)=
\delta^{ab}\left(\delta_{\mu\nu}-\frac{q_{\mu}q_{\nu}}{q^2}\right)
\frac{Z(q^2)}{q^2} \; ,
\end{equation}
should behave in the infrared as 
$Z(q^2) \propto (q^2)^{\kappa_D}$ with $\kappa_D > 1$.

\item The ghost propagator ought to be more singular at $q \to 0$ than a free 
scalar one~\cite{Zwanziger:1991ac,Suman:1995zg}. 
This is the {\it horizon condition}. The ghost dressing function, defined through
\begin{equation}
G^{ab}(q) = \delta^{ab} \frac{J(q^2)}{q^2} \; ,
\end{equation}
should behave in the infrared as 
$J(q^2) \propto (q^2)^{\kappa_G}$ with $\kappa_G < 0$.

\item Positivity of the spectral function is expected~\cite{Zwanziger:1991ac} 
to be violated for the gluon propagator,
meaning that the weight function $\rho(m^2)$ in the K\"allen-Lehmann representation
\begin{equation}
D(q^2) = \int_{0}^{\infty} \frac{dm^2~\rho(m^2)}{q^2 + m^2}\; \; , \qquad
D(t,{\vec q}=0) = \int_{0}^{\infty}dm^2~\rho(m^2)~e^{-mt} 
\end{equation}
would no longer be $\rho(m^2) \geq 0$ for all $m^2$ .

\item The Kugo-Ojima (KO) confinement criterion~\cite{Kugo:1979gm} 
is formulated in terms of the ghost 
propagator $G^{ab}_{xy}=\langle c^a_x \bar{c}^b_y \rangle$. 
One defines $u^{ab}(q^2)$ by 
\begin{equation}
\int d^4x e^{iq(x-y)} \langle (D^{ac}_{\mu}c^e)_x (f^{bcd}A^d_{\nu}\bar{c}^c)_y 
\rangle = \left(\delta_{\mu\nu}-\frac{q_{\mu} q_{\nu}}{q^2} \right) u^{ab}(q^2)
\end{equation}
and requires that $u^{ab}(q^2) \to - \delta^{ab}$ in the limit $q^2 \to 0$.
This guarantees the absence of colored asymptotic states. 
The criterion was derived from the so-called quartet mechanism
within the BRST quantization of Yang-Mills theory.
\end{itemize}

The practical request from the side of hadron physics has stimulated
non-perturbative studies in the continuum of the gluon and ghost propagator 
that have started ten years 
ago~\cite{Smekal:1997is,Smekal:1998vx,Alkofer:2000wg}. 
The authors were seeking for solutions of the hierarchy of Dyson-Schwinger 
equations (DSE) (coupled for both propagators) adopting some truncations. 
In this approach infinite volume presents no problem. More recently one has 
learned how to solve the DSE in a compactified space, on the 
4-torus~\cite{Fischer:2005ui,Fischer:2007pf}. 
The lessons from DSE, for infinite and compactified space, provide a 
framework to discuss the status of the lattice calculations. It helps to 
orient oneself on the ``long march'' to the infinite-volume limit. 

The particular value of {\it lattice calculations} at first consists 
in their ability to {\it control} the assumptions and truncations made in 
the DSE approach. At second and even more interesting as I find, they are 
possible to assess the importance of special confining field excitations 
(monopoles and vortices, dyons and calorons) and/or external conditions 
on the functional form of the propagators. At third, from the beginning of 
the lattice studies it was clear that the Gribov ambiguity would present a 
hard problem. 

Thus, it is left to the {\it lattice studies} to elucidate the open theoretical 
problems how to deal with it. If the lattice discretization  
{\it is the definition} of QCD in the non-perturbative regime, different 
prescriptions how to take into account the Gribov problem could lead to different 
versions of QCD requiring verification.

The vanishing (divergence) of the gluon (ghost) propagators can be traced back to
the restriction inside the so-called {\it Gribov region} $\Omega$ 
of the gauge field representants $A_{\mu}$ (transverse gauge copies) that are 
contributing to the path integral.
This is the region where the Faddeev-Popov operator ${\cal M}$ is positive.
The problem are more than one of such copies.
In the infinite-volume limit the tendency emerges that the most important 
configurations concentrate at the boundary, the Gribov horizon, 
such that small non-trivial eigenvalues 
of ${\cal M}$ accumulate close to zero with a finite density. This is the 
{\it Gribov-Zwanziger confinement scenario}~\cite{Zwanziger:1991ac}. 
The infrared exponents 
$\kappa_D = 2 \kappa \approx 1.2$ and $\kappa_G = -\kappa \approx -0.595$ 
(constrained to $1/2 < \kappa < 1$~\cite{Lerche:2002ep}) have been obtained 
both by the DSE approach~\cite{Watson:2001yv,Lerche:2002ep} and by stochastic 
quantization~\cite{Zwanziger:2001kw,Zwanziger:2002ia}. A consequence of the 
interrelation between both infrared exponents is an infrared fix-point of the 
strong coupling, $\alpha_s(0)=8.915/N_c$ .

The positivity violation was noticed very 
soon~\cite{Mandula:1987rh,Bernard:1992hy,Marenzoni:1993td}
in lattice simulations, when the ``local mass'' 
$m_{eff}(t) = - d \log D(t,{\vec q}=0)/dt > 0$ was 
found to {\it increase} with increasing $t$. For a physical particle in the asymptotic
Hilbert space, the effective mass $m_{eff}(t)$ approaches the actual mass from above.

\section{The lattice framework}

Lattice gauge theory is formulated in a way that circumvents the choice 
of a gauge. Apart from our task (to calculate Green's functions) there are 
many other contexts in which fixing the gauge is necessary or useful.
Gauge-fixing usually becomes a very time-consuming part of such calculations 
and deserves particular attention. The procedure of such a calculation is as 
follows: An ensemble of gauge configurations $\{U\}$ is generated with one's 
favorite action using the Monte Carlo (MC) method, either without (``quenched'') 
or with the back-reaction of (``dynamical'') quarks through the fermion determinant 
taken into account. In the quenched approximation one has just a gluonic inverse 
``bare coupling'' $\beta$, and the lattice spacing $a$ is a function of it,
$a(\beta)$, that can be defined by putting the string tension 
$\sigma=a^{-2} \sigma_L(\beta)$ equal to some physical value. Up to a global scale, 
the renormalization of the gluon propagator (matching the propagators measured  
at different $\beta$) is an independent way to define the running lattice scale 
$a(\beta)$. 

The vector potential needs to be extracted from the basic transporters 
(``links'') 
as ${\cal A}_{x+\hat{\mu}/2.\mu}
=\left( U_{x\mu} - U^{\dagger}_{x,\mu} \right)_{\rm traceless}/(2iag)$.
In order to implement the gauge in question, every gauge configuration 
$\{U\}$ has to be gauge-transformed $U_{x,\mu} \to U^{g}_{x,\mu} 
= g_x U_{x,\mu} g^{\dagger}_{x+\hat{\mu}}$
by a suitable $\{g\}$. 

\noindent
For example, for the Landau gauge an extremization
\begin{equation}
F_{U}[g]=\frac{1}{N_c} \sum_{x,\mu} 
\rm{Re}~{\rm Tr}~g_x U_{x,\mu} g^{\dagger}_{x+\mu} \to {\rm Max}
\end{equation}
with respect to $\{g\}$ solves the problem. A {\it local maximum} is found when 
\begin{equation}
(\partial_{\mu} {\cal A}^{g}_{\mu})_x 
= \sum_{\mu} \left( {\cal A}^{g}_{x+\hat{\mu}/2,\mu} -
                    {\cal A}^{g}_{x-\hat{\mu}/2,\mu} \right) = 0 \; 
\end{equation}
(the transversality condition) is satisfied with high precision.
This defines the recommended stopping criterion for the various iterative 
gauge-fixing methods.
Having found a local maximum $\{g\}$, 
{\it for any infinitesimal} $\tilde{g}$, one has 
$F_{U^{g}}[\tilde{g}] < F_{U^{g}}[1]$.  
For the {\it absolute maximum} $\{g\}$, this should hold {\it for all} $\tilde{g}$. 
Thus, the gauge-fixing problem has been put into the form of a disordered spin 
system. The search for the (classical) ground state of a spin glass is known to be 
a {\it non-polynomially hard} problem. 

If extracting physics would depend on the ability to find the absolute maximum
one had to stop here. In this case the measure is said to be restricted to the 
so-called {\it fundamental modular region} $\Lambda$.
It is possible, however, to go a bit further and to
investigate the convergence of gauge-variant observables with an increasing 
number $n_{\rm copy}$ of Gribov copies. 
A sequence of replica ensembles labelled by $n_{\rm copy}$ is recursively created
(with $n_{\rm copy}=0$ denoting to the original MC ensemble). Each time one steps 
from $n_{\rm copy} \to n_{\rm copy} +1$, for each MC configuration a new
gauge-fixing attempt is made starting from a random gauge transformation. 
If a better representant of the original MC configuration is found, it replaces
the ``previously best'' copy, such that the $n_{\rm copy}$-th ensemble is an 
ensemble of ``currently best'' copies after $n_{\rm copy}$ attempts.

On the other side, Zwanziger~\cite{Zwanziger:2003cf} gave arguments that in the 
infinite-volume limit an average over all gauge-fixed copies in the Gribov 
region would be the physically correct prescription. This would make the search
for ever better copies obsolete, and it would be just a question of statistics
how many gauge copies of one MC configuration are evaluated.

In any case, for the present lattice sizes it is important to assess the 
gauge copy dependence of the propagators. Following the ``best copy vs. first 
copy'' strategy, one sees that the dependence is stronger at small momenta 
and becomes indeed weaker with increasing volume.

In order to do the maximization, methods like overrelaxation (OR), 
Fourier accelerated gauge-fixing (FA) and simulated annealing (SA) are practically 
in use. The latter~\cite{Bali:1996dm} is a quasi-equilibrium MC process with a 
probability distribution $\propto \exp(F_U[g]/T)$. Annealing means that the
temperature is guided from $T_{\rm max}$ down to $T_{\rm min}$.
The idea is that OR following the SA (until the transversality is satisfied)
finds the finally gauge-fixed copy with only few iterations within one basin of 
attraction. Therefore, improvement of the gauge-fixing is not mainly aiming
to accelerate the relaxation but to increase the yield of ``good'' gauge-fixed 
copies, as close as possible to the best copy. Given this objective, 
SA strategies become superior on large lattices~\cite{Schemel:2007xx} also in 
terms of computing time.

The gluon propagator is defined immediately in momentum space by correlating
Fourier transforms $\tilde{{\cal A}^{g}}$ of the ${\cal A}^{g}$ field,
\begin{equation}
D^{ab}_{\mu\nu}(q) = \langle \tilde{{\cal A}^{g}}^a_{\mu}(k)
                             \tilde{{\cal A}^{g}}^b_{\nu}(-k) \rangle \; ,
\end{equation}
where the finite lattice Fourier transform is calculated for integers 
$k_{\mu} \in (-L_{\mu}/2,L_{\mu}/2]$. The momentum vector
$q_{\mu}(k_{\mu})=(2/a) \sin\left( \pi k_{\mu}/L_{\mu} \right)$
is associated to them. If the gluon propagator is to be calculated for many momenta, 
use of fast Fourier transformation is necessary. 

The ghost field is not a $c$-number field in the memory, such that the ghost 
propagator, similar to a quark propagator, must be obtained by inversion 
of the Faddeev-Popov operator
\begin{equation}
{\cal M}^{ab}_{xy}(U)=\sum_{\mu}\left( A^{ab}_{x,\mu}(U)\delta_{xy} 
                              - B^{ab}_{x,\mu}(U)\delta_{x+\hat{\mu},y} 
			      - C^{ab}_{x,\mu}(U)\delta_{x-\hat{\mu},y} \right) 
			              \; ,
\end{equation} 
with ${\cal M}^{ab}_{xy} \to - \delta^{ab} \Delta_{xy}$ for $U_{x,\mu} \to 1$.
The matrices $A$, $B$ and $C$ are defined in terms of (gauge fixed) links as
\begin{eqnarray}
A^{ab}_{x,\mu}  &=& \rm{Re}~{\rm Tr}\left[ {T^a,T^b} 
\left( U_{x,\mu} + U_{x-\hat{\mu},\mu} \right) \right] \; ,  \nonumber  \\
B^{ab}_{x,\mu}  &=&  2~\rm{Re}~{\rm Tr}\left[ T^b T^a U_{x,\mu} \right] \; \\
C^{ab}_{x,\mu}  &=&  2~\rm{Re}~{\rm Tr}\left[ T^a T^b U_{x-\hat{\mu},\mu} \right] \; . 
                                                                       \nonumber
\end{eqnarray}
In momentum space the propagator is obtained by inverting ${\cal M}$ 
on a plane wave 
source (for $k\ne(0,0,0,0)$)
\begin{equation}
\sum_{b,y} {\cal M}^{ab}_{xy} \phi^{b(c)}_y = \psi^{a(c)}_x  
= \delta^{ac} e^{2 \pi i k \cdot x} \; ,
\end{equation}
giving 
\begin{equation}
G^{ab}(q) = \frac{1}{V (N_c^2-1)} \sum_{c} \sum_{x} \psi^{a(c)*}_x \phi^{b(c)}_x \; .
\end{equation}
For the inversion the conjugate gradient algorithm is used. For preconditioning
one uses the simple (not the covariant !) Laplacian.  

If one defines the strong coupling $\alpha_s$ through the 
ghost-gluon vertex, then, knowing the (renormalized) dressing functions $Z_R$ 
and $J_R$ and assuming for the vertex renormalization constant $Z_1(q^2) \approx 1$,
one obtains~\cite{Taylor:1971ff}
in the MOM-scheme the running coupling as follows 
\begin{equation}
\alpha_R(p^2) = \alpha_R(\mu^2)~Z_R(p^2,\mu^2)~\left[J_R(p^2,\mu^2)\right]^2 \; .
\label{eq:coupling}
\end{equation}

\section{Some lattice results}

The lattice calculations should give an answer to the following questions:
\begin{itemize}
\item Do the propagators show the infrared behavior proposed by DSE ?
\item What is the infrared limit of the MOM-scheme coupling $\alpha_s(q^2)$ ?
\item What is the impact of Gribov copies on the propagators ?
\item How fast is the infinite-volume limit reached ?
\item Which propagators are modified by ``unquenching'' ? 
\item How are the other confinement criteria fulfilled ?
\item How do Faddeev-Popov eigenvalues and eigenmodes behave ? 
\item Is the ghost propagator in the infrared dominated by the lowest 
eigenmodes of ${\cal M}$ ?
\end{itemize}
\noindent Finally, one might ask:
\begin{itemize}
\item Are there modified gauge-fixing conditions, equivalent to the
common ones in the infinite-volume limit, that are advantageous for convergence 
to the infinite volume limit and/or less vulnerable to discretization effects ?
\end{itemize}

I will present some answers in the following.
Our studies have included quenched $SU(3)$ QCD on lattices from 
$12^4$ to $72^4$ generated 
with Wilson gauge action at $\beta=5.7$, $5.8$, $6.0$ and $6.2$.
The full QCD configurations kindly provided by the QCDSF collaboration are 
$16^3\times32$ and $24^3\times48$ lattices created with Wilson gauge action 
at $\beta=5.29$ and $5.25$ and $N_f=2$ clover-improved Wilson fermions of 
varying mass 
($\kappa=0.135$ ... $0.13575$). The last question of an improved gauge-fixing
was recently investigated in quenched $SU(2)$ gauge 
theory~\cite{Bogolubsky:2007bw} where the consequences of enlarging the set of 
admissible gauge transformations by global $Z(N)$ flips 
(proposed in \cite{Bogolubsky:2005wf}) were further examined.

In Fig. 1a we show the gluon dressing function for quenched 
QCD~\cite{Sternbeck:2005tk,Sternbeck:2005re}.
Characteristic is the intermediate bump of the dressing function.
The exact form of the dressing function is {\it not described} by the DSE,
which pretend to describe only the infrared and ultraviolet behavior.
In particular the bump is underestimated. Fig. 2a shows that this enhancement 
becomes partly (30 \%) depressed by the back-reaction of dynamical 
quarks~\cite{Ilgenfritz:2006gp}. The same
is observed for dynamical configurations of the MILC collaboration
in \cite{Bowman:2007du}. In view of the difficulties to determine
the infrared exponent $\kappa_D$ (see below) it is premature to speak
about the dynamical-quark effect on $\kappa_D$. Since the main effect 
is not in the infrared behavior, the change could be considered irrelevant
for the confinement problem.  Indeed, breakdown 
of {\it gluon confinement} is not realistic in the real world with 
dynamical quarks. In contrast to that, it is known that dynamical quarks 
indeed change the 
confinement property of static quarks (``string breaking''~\cite{Bali:2005bg}). 
In the quenched $SU(2)$ theory the so-called ``infrared bump'' (sitting, however
in fact at 1 GeV !) is entirely the result of the presence of P-vortices as 
confining agents seen in Maximal Center Gauge (MCG) and projection. The enhancement 
By the same operation confinement~\cite{Boyko:2006ic}, topological charges and chiral 
symmetry breaking~\cite{Bornyakov:2007fz} are destroyed.
A natural conjecture is that dynamical quarks to some extent suppress 
P-vortices. This hypothesis deserves closer investigation.
That the opposite effect of unquenching is observed for the 
density of monopoles~\cite{Bornyakov:2003vx} can be explained that there is an 
``inert'' component of monopoles~\cite{Boyko:2006ic} not related to P-vortices. 

\begin{figure}[ht]
\begin{center}
\epsfig{file=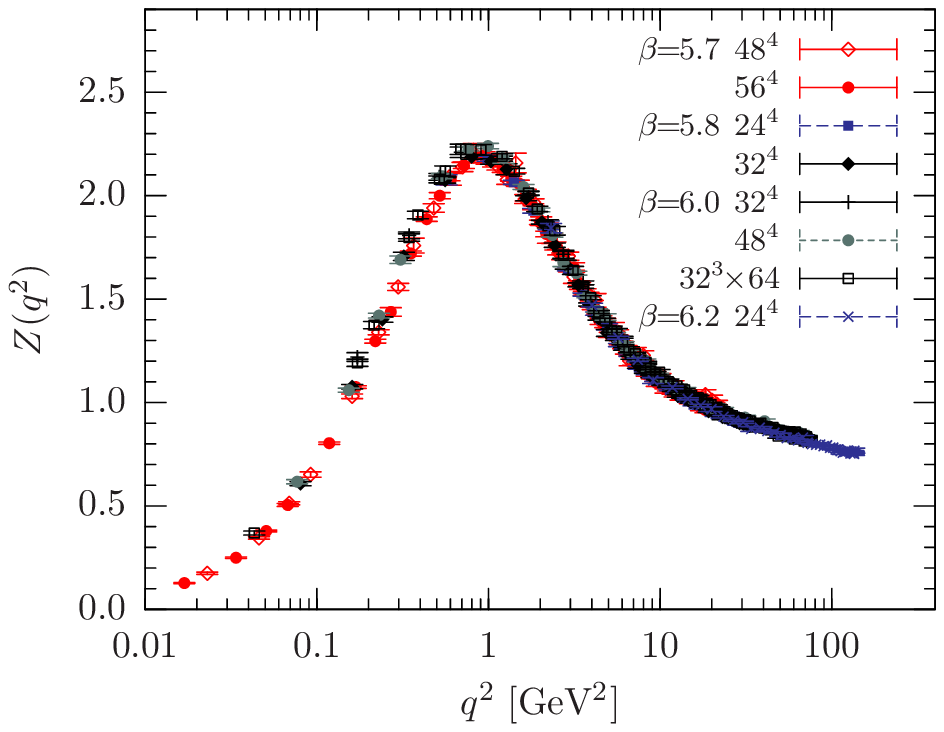,angle=0,width=9cm} \\
          (a)  \\
\vspace{1cm}
\epsfig{file=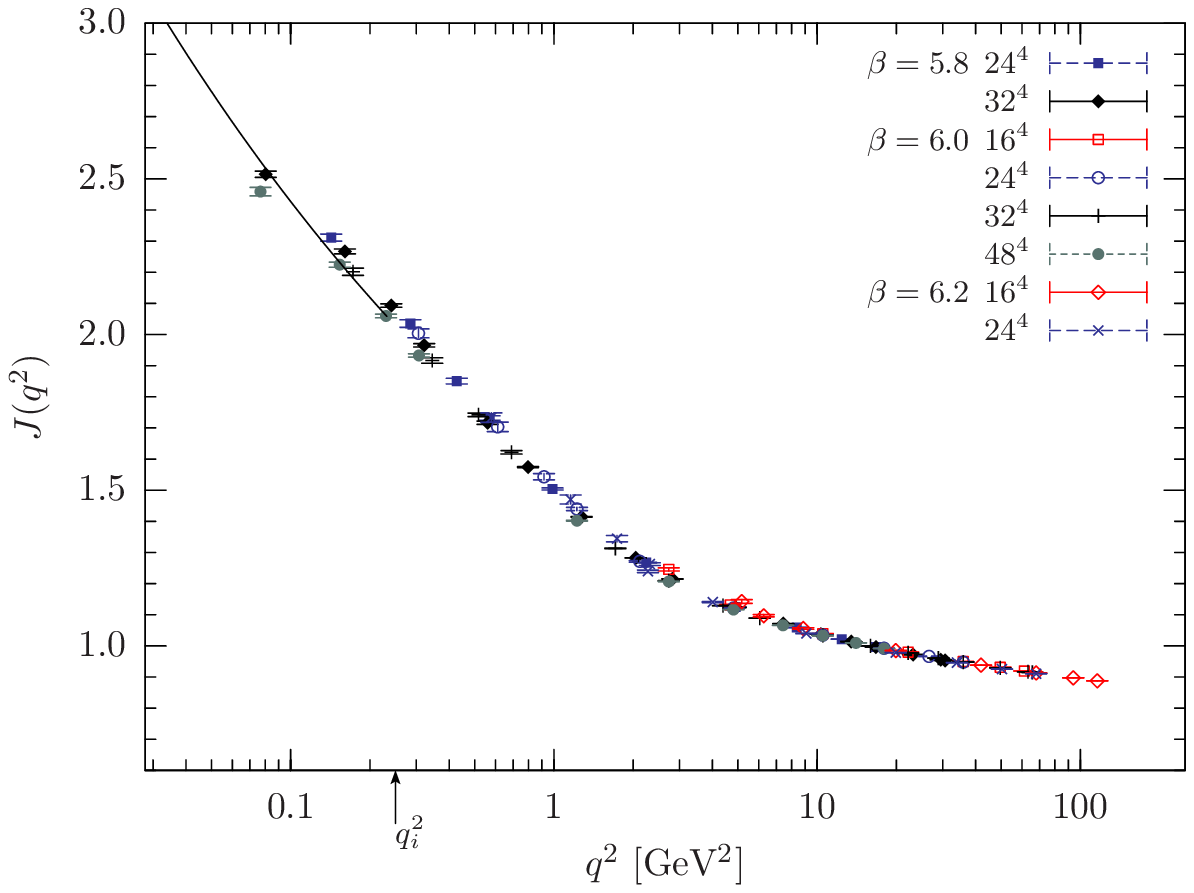,angle=0,width=9cm}\\
          (b)
\end{center}
\caption{The gluon dressing function (a) and the ghost dressing function (b)
for quenched QCD. Data from various lattice sizes and $\beta$-values are seen
matching on one curve.
The little $q_i^2$ marks a momentum range $q^2 < q_i^2$ where a power fit for
$\kappa$ has been attempted. Both propagators give a $\kappa \approx 0.2$ .} 
\end{figure}

Fig. 1b presents the ghost dressing function~\cite{Sternbeck:2005tk,Sternbeck:2005re}
for the quenched theory. The behavior in the infrared is opposite to the gluon
dressing function and not incompatible with being divergent.
In Fig. 2b one sees that unquenching~\cite{Ilgenfritz:2006gp} has no dramatic effect 
on the ghost propagator, except for the smallest momenta accessible, where also 
a splitting according to the quark masses (see the legend) becomes visible.  

The infrared increase of the ghost structure function in the quenched theory 
is obvious, but the fitting of an infrared exponent does not give the 
expected $\kappa$. For $SU(2)$ gauge theory it is known~\cite{Gattnar:2004bf}  
that the removal of P-vortices leads to a global change of the ghost dressing 
function $J(q^2)$ which becomes almost constant. 
One can say that the global (not only infrared) behavior of the ghost dressing 
function is the closest relative to the confinement of quarks. 
Since vortex removal also removes all non-perturbative 
attributes~\cite{deForcrand:1999ms} 
(percolating monopole trajectories, string tension, chiral condensate and the
topological charge~\cite{Bornyakov:2007fz}, it is very likely that the original 
divergence of the ghost 
propagator like $1/(q^2)^{1+\kappa}$ is mainly a result of the topological 
structure leading to an enhanced density of low-lying 
eigenvalues of ${\cal M}$ as demonstrated for MC~\cite{Greensite:2004ur} and
model configurations~\cite{Maas:2006ss}.
We have found, however, that the direct correspondence between the ghost 
propagator at lowest momenta and the lowest-lying Faddeev-Popov eigenmodes 
is rather weak~\cite{Sternbeck:2005vs}. 
The effect of dynamical quarks is not as strong as vortex removal, but it
might be caused indirectly via the gradual suppression (or pairing) of topological 
objects by light sea quarks, too. The quark mass dependence of the effect might 
be related to the stronger string breaking induced by lighter sea quarks. 
 
Our data suggest that both in the quenched and the dynamical case 
there are apparently no finite-volume effects on the ghost propagator. 
This will be made more precise later.

Figs. 1 and 2 for the gluon propagator come from a study where the Gribov 
ambiguity was ignored. Only one gauge-fixed copy (``first copy'') was evaluated. 
I should remark that this procedure is equivalent to the prescription of averaging 
over all gauge-fixed copies within the Gribov region 
(justified in \cite{Zwanziger:2003cf}) for a given MC configuration.

\begin{figure}[ht]
\begin{center}
\epsfig{file=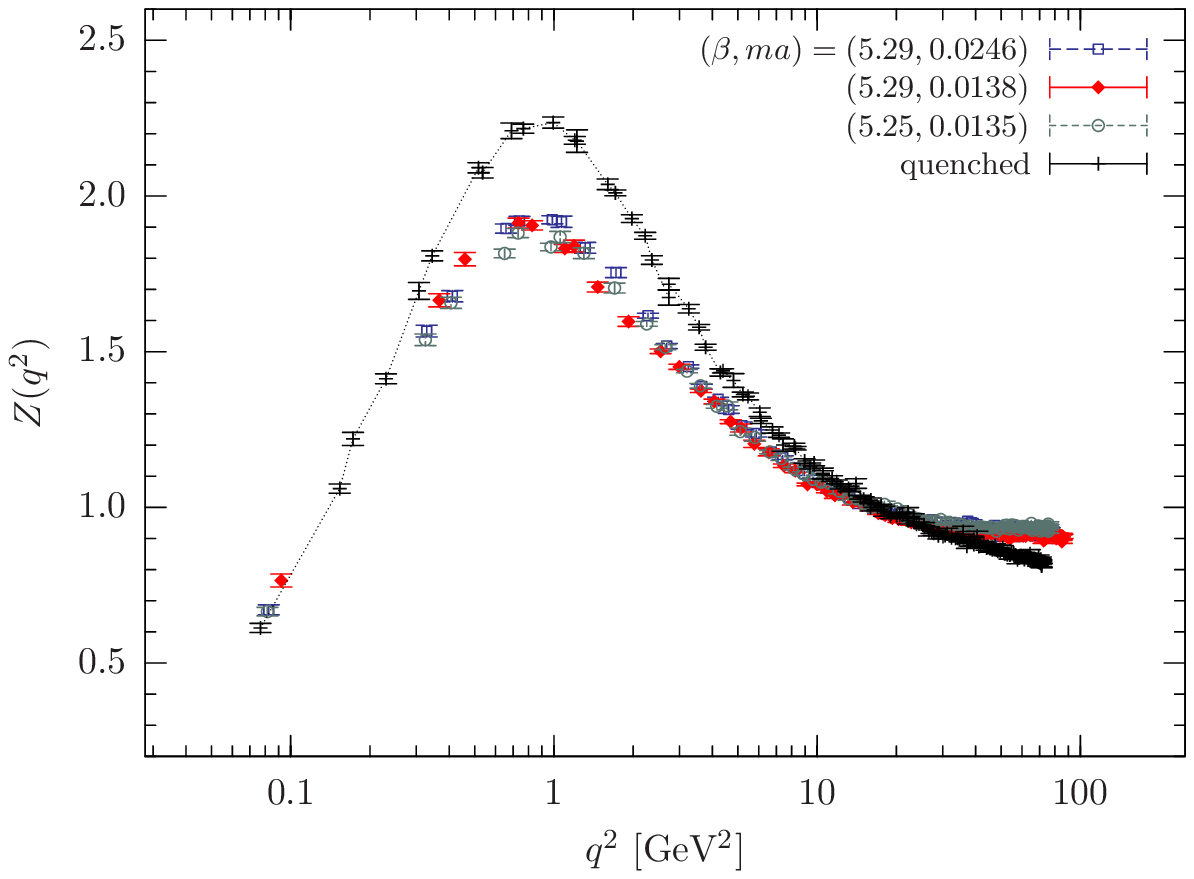,angle=0,width=9cm} \\
          (a)  \\
\vspace{1cm}
\epsfig{file=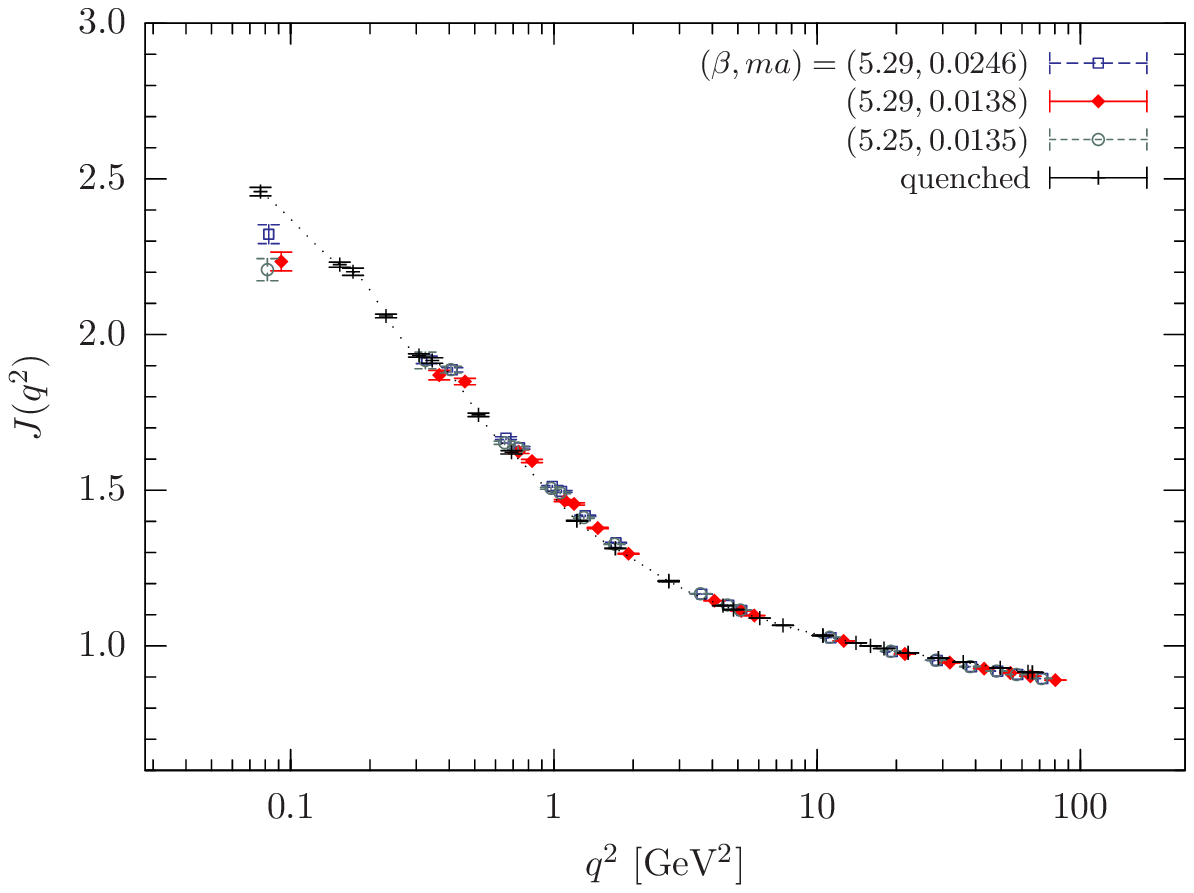,angle=0,width=9cm}\\
          (b)
\end{center}
\caption{The effect of dynamical quarks 
(a) on the gluon propagator that becomes depressed in the intermediate 
momentum range around $O(1 {\rm~GeV})$ ;
(b) on the ghost propagator that becomes depressed only in the  
infrared region.}
\end{figure}

In order to demonstrate that the propagators are all vulnerable to the Gribov
ambiguity, but to a different extent, in Fig. 3 we present (for smaller lattices) 
the effect of the Gribov ambiguity on the gluon and ghost propagator (for the 
quenched case)~\cite{Sternbeck:2005tk,Sternbeck:2005re}.
In the subpanels (a) and (b) the ratio of the dressing functions 
calculated in two different ensembles is shown. 
The ``fc'' ensemble is the ensemble of (arbitrary) first gauge-fixed copies
for each MC configuration, ``bc'' is the ensemble of the best copies after
$n_c=20$ to $30$ gauge-fixing attempts. In the case of the gluon propagator in 
Fig. 3a we see a 
relatively broad band of ``Gribov noise'' that does not show a distinct
momentum or volume tendency. On the other side, for the ghost dressing 
function in Fig. 3b a relatively sharp effect of overestimation for the 
first copy is seen that becomes stronger towards smaller momenta. 
The effect is slightly suppressed with increasing physical volume 
(see the data points for the lowest $\beta=5.8$), an observation that
can be an early hint towards the weakening of the Gribov copy effect at 
very large volumes. For smaller lattices, however, the ghost propagator will 
be overestimated at the smallest momenta if there is no systematical search
for better Gribov copies, i.e. if one averages over Gribov copies.

\begin{figure}[ht]
\begin{center}
\epsfig{file=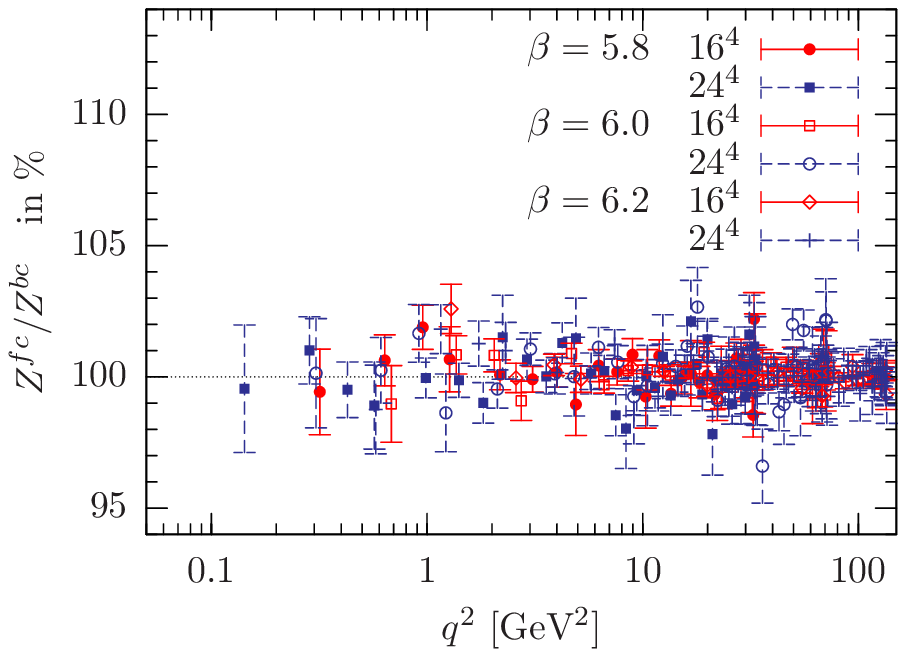,angle=0,width=9cm} \\
          (a)  \\
\vspace{1cm}
\epsfig{file=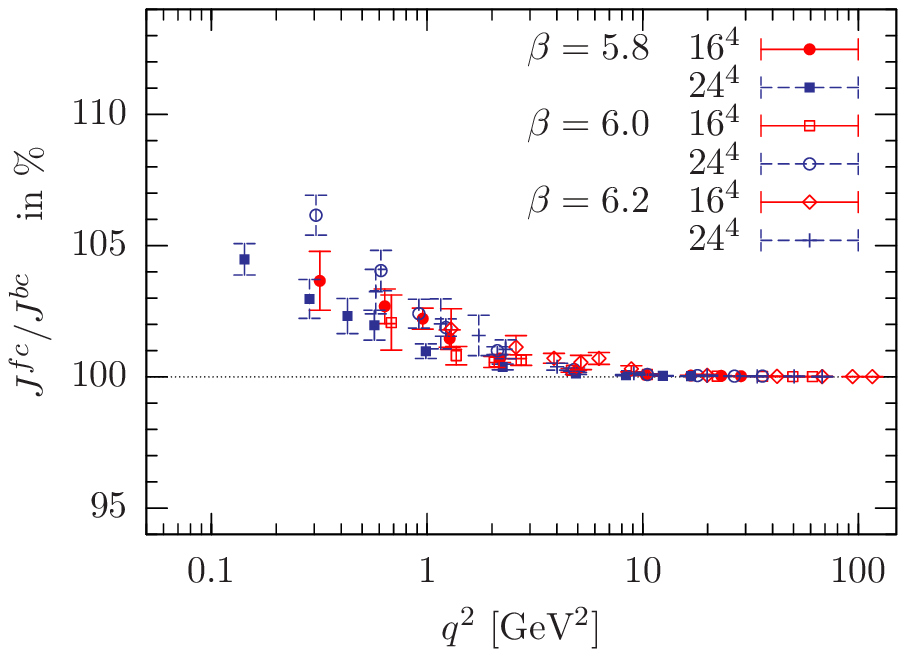,angle=0,width=9cm}\\
          (b)
\end{center}
\caption{The ratio between first and best gauge copies 
used for calculating the dressing function  
(a) for the gluon propagator (``Gribov noise''), 
(b) for the ghost propagator where one sees a systematic Gribov copy effect
becoming weaker with increasing physical volume.} 
\end{figure}

There are not enough data yet in the region of small enough momenta 
to get stable fits of the infrared exponents. If one attempts this, $\kappa$
is found too small. In order to anticipate whether the gluon propagator 
finally may turn to zero in the limit $q^2 \to 0$, one looks at 
the propagator instead of the dressing function. Fig. 4 shows 
data for the gluon and the ghost propagator on a $64^4$ lattice at $\beta=5.7$
obtained at the MVS-15000BM of the Joint Supercomputer Center (JSCC) Moscow.
The gluon propagator in Fig. 4a shows at least a kind of plateau.
The leftmost data point represents the gluon propagator at zero momentum,
$D(0)$. The decreasing tendency of $D(0)$ with the lattice volume (not shown here) 
suggests that the propagator function $D(p)$ also cannot be taken as the 
infinite-volume limit. More recent data (on a lattice $80^4$) presented at Lattice 2007
~\cite{Bogolubsky_poster} indicate that the plateau extends to $|q|$ below 100 MeV.
The ghost propagator in Fig. 4b (shown in a log-log-plot) suggests already
something close to a power law, but the corresponding $\kappa$ comes also 
too small compared with the preferred $\kappa=0.595$. 

\begin{figure}[ht]
\begin{center}
\epsfig{file=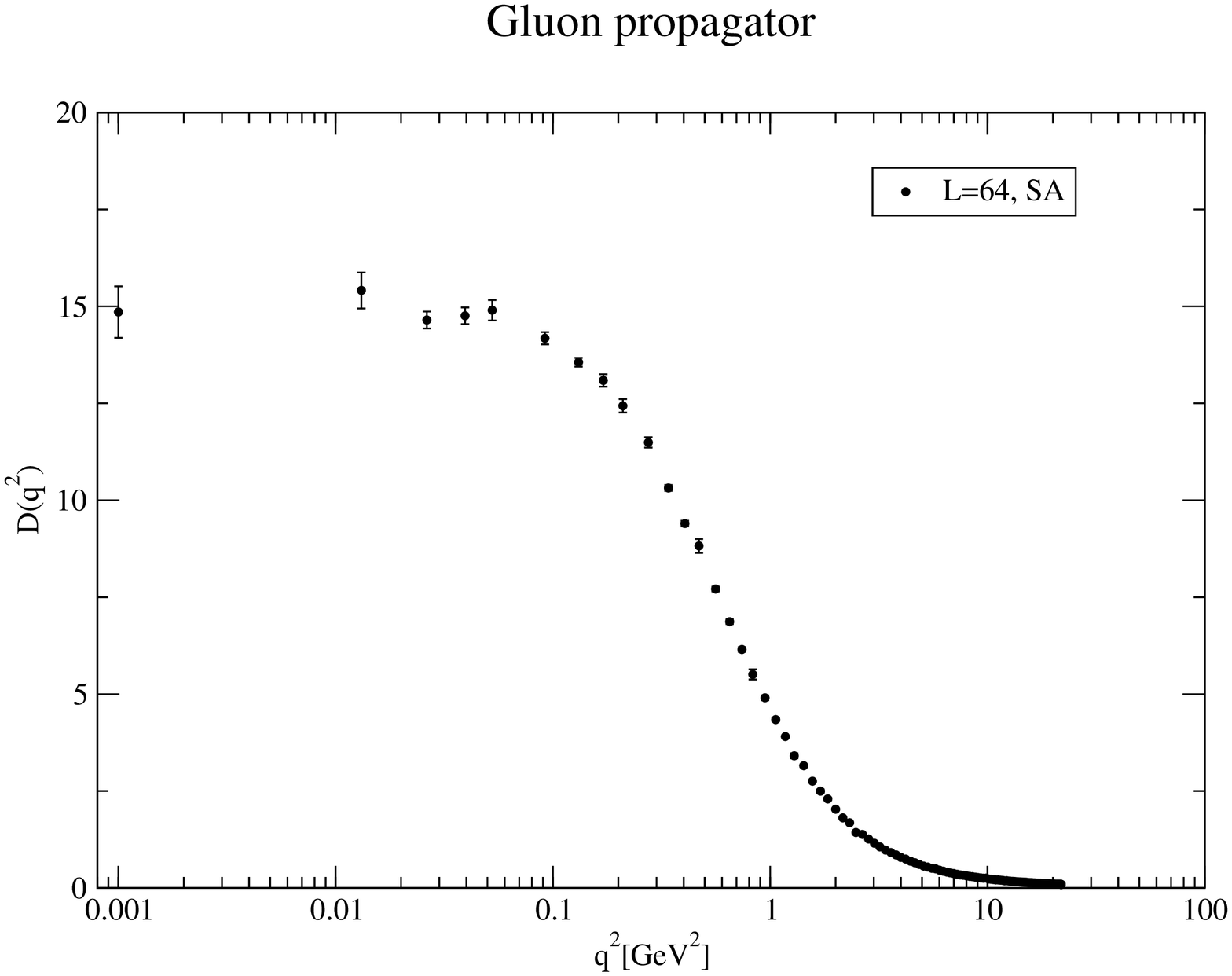,angle=270,width=9cm} \\
          (a)  \\
\epsfig{file=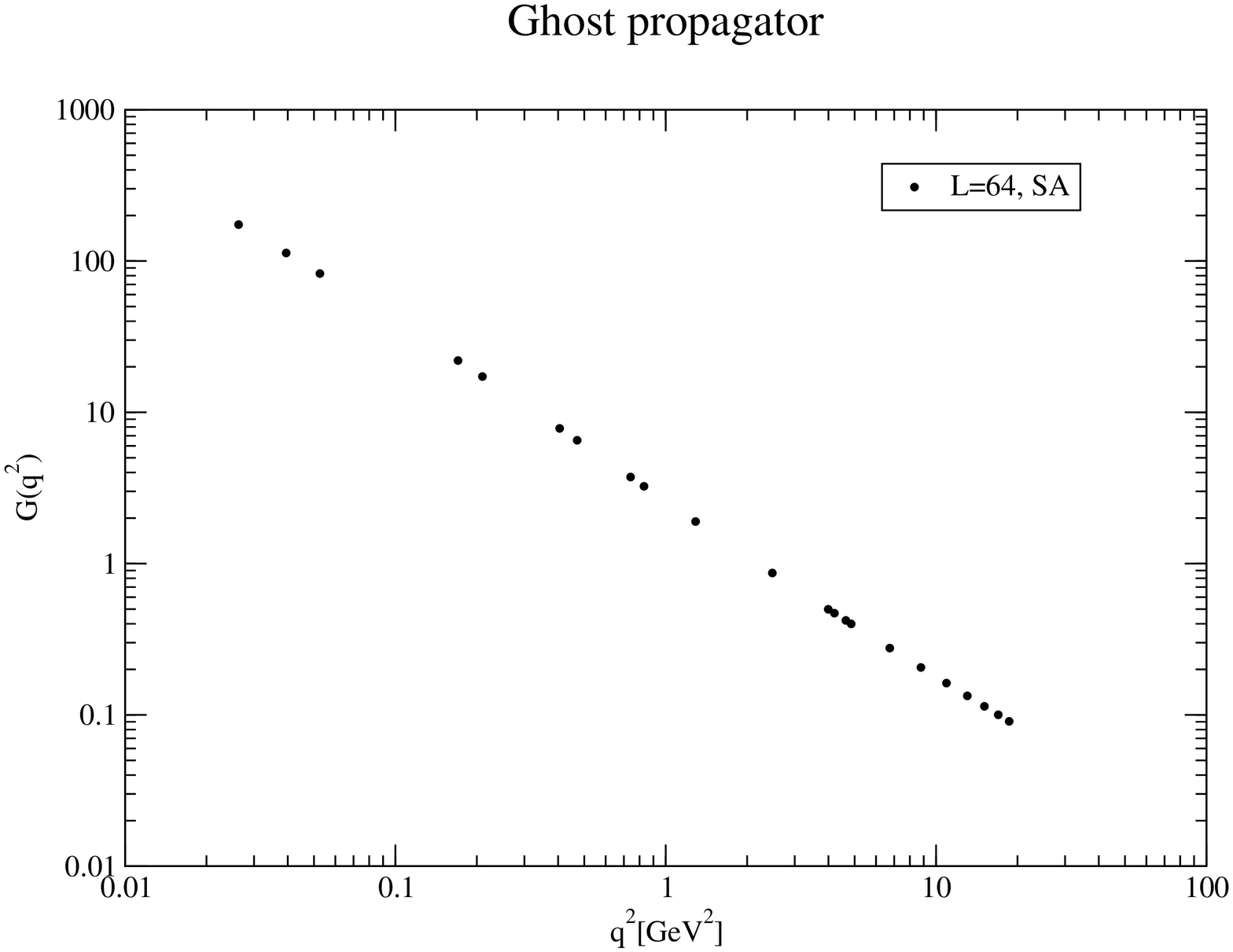,angle=270,width=9cm}\\
          (b)
\end{center}
\caption{(a) The gluon propagator, (b) the ghost propagator, measured  
on the $64^4$ lattice at $\beta=5.7$ .
Notice that the leftmost data point of the gluon propagator actually 
refers to zero momentum, $D(0)$. So far there is no indication that 
$D(p) \to 0$ for $p \to 0$ . } 
\end{figure}

That means that the now accessible momentum range is probably still pre-asymptotic.
DSE results anticipating the approach to the infinite-volume limit indicate
how far lattice calculations are from seeing the asymptotic behavior.
The DSE have been formulated and solved on a 
finite torus~\cite{Fischer:2005ui,Fischer:2007pf}, 
and an interesting pattern of 
finite-volume deviations for the calculated propagators has been found
and compared with our lattice data (see Figs. 5a and 5b taken from 
\cite{Fischer:2007pf}).  

\begin{figure}[ht]
\begin{center}
\epsfig{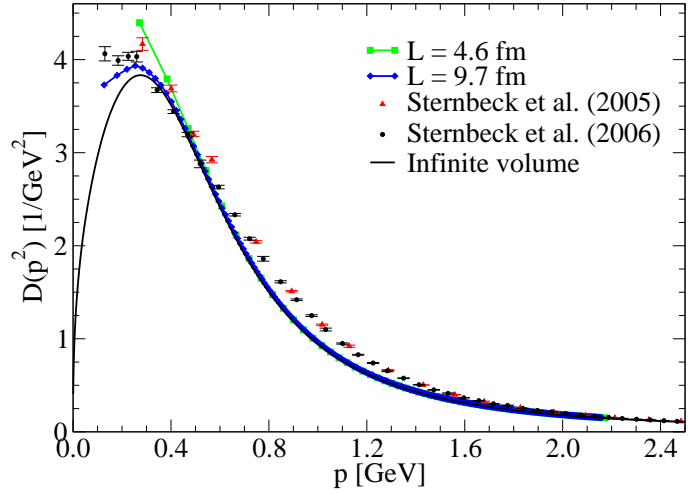} \\
          (a)  \\
\vspace{1cm}
\epsfig{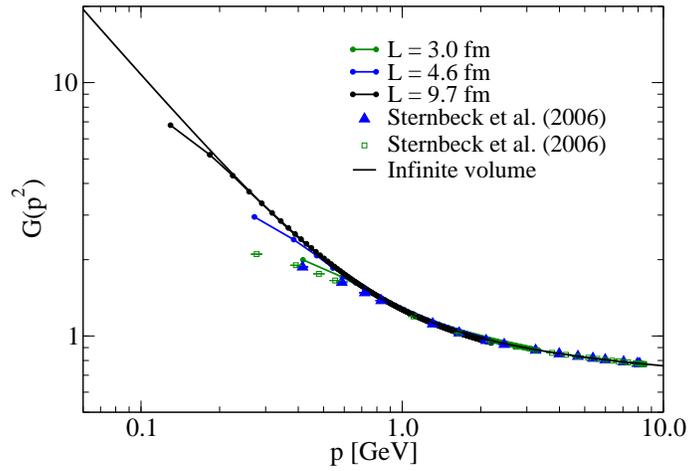} \\
          (b)
\end{center}
\caption{The propagators from DSE for a few volumes and for
infinite volume shown together with our 
lattice data of \cite{Sternbeck:2005tk,Sternbeck:2006cg},
(a) for the gluon propagator, where the lattice data, being tangents to the 
infinite-volume curve, are strongly deviating upward;
(b) for the ghost propagator, where the lattice data, being tangents to the 
infinite-volume curve, are deviating downward with a pre-asymptotic slope.
(Figures taken from \cite{Fischer:2007pf})}
\end{figure}

For the gluon propagator the approach to the infinite-volume curve 
is from above, with an enormous overshooting towards the lowest momentum for
any given lattice volume (see Fig. 5a). 
For the ghost propagator the approach is from below and less dramatic. 
This is shown in Fig. 5b. The insufficient slope $\kappa$ in the log-log-plot 
of the ghost propagator in Fig. 4b is well explained by this type of 
finite volume effect. 

Fig. 6a shows the DSE result for the running coupling with the volume
dependence induced by the volume dependence of the gluon and ghost
propagators. This makes clear that it is illusory to see the running coupling
approaching the infrared fix-point before lattices reach a linear size
$L=O(15 {\rm~fm})$. We have checked~\cite{Sternbeck:2005re,Ilgenfritz:2006he} 
on the lattice the assumed $q^2$ independence
of the ghost-ghost-gluon vertex renormalization constant, a tacit assumption
in deriving Eq. (\ref{eq:coupling}).  
Fig. 7b shows the result of our calculation of the gluon and 
ghost dressing functions, giving the running coupling~\cite{Sternbeck:2005re}. 
The volume is just large enough to reveal the turn-over to an apparently
decreasing behavior of coupling with $q^2 \to 0$. But this has nothing to do
with the true asymptotic behavior. In the light of this observation, 
the optimism of {\it having seen} already 
the approach to the fix-point~\cite{Bloch:2003sk} seems to be premature.

\begin{figure}
\begin{center}
\epsfig{file=Figures/fischer_fig4.eps,angle=0,width=9cm} \\
          (a)  \\
\vspace{1cm}
\epsfig{file=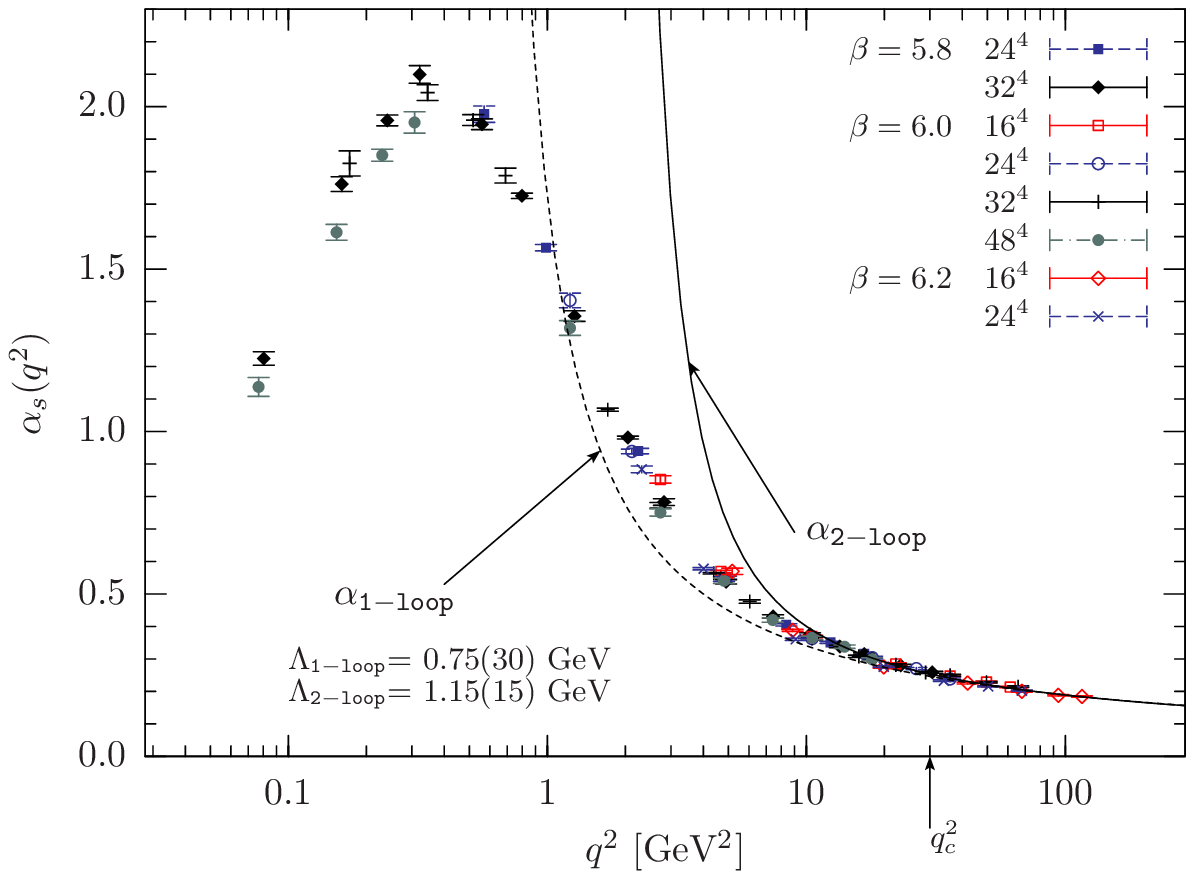,angle=0,width=9cm}\\
          (b)
\end{center}
\caption{The MOM-scheme running coupling constant $\alpha_s(q^2)$ :
(a) from the DSE approach for different box sizes compared with the infinite 
volume limit (taken from \cite{Fischer:2007pf}),
(b) from our quenched lattice calculations. 
The little $q_c^2$ indicates the end of a fit range $q^2 > q_c^2$ where
the 1- and 2-loop running coupling has been fitted. Notice the weak volume 
dependence on the right of the peak and the beginning splitting on the left 
of the peak between the largest volumes.}
\end{figure}

The violation of positivity and the very slow approach to the KO confinement 
criterion have been presented at Lattice 2006~\cite{Sternbeck:2006cg}.
Recently, the Adelaide group~\cite{Bowman:2007du} has discussed violation of 
positivity together with scaling and the effect of dynamical quarks in much 
more detail for lattice ensembles provided by the MILC collaboration.

\vspace{-0.3cm}
\section{Summary}

Various effects on the gluon and ghost propagators have already been
studied for quenched QCD, and the effect of dynamical quarks and Gribov
copies has been added by our investigations. The infrared exponents 
characteristic for the way how ``ghosts manage to confine gluons'' are 
still elusive.
The infrared asymptotic region in momenta (volumes) is not yet reached.
There are three extrapolations needed before lattice QCD can be applied to 
the real world: (a) to take the continuum limit, (b) to control the chiral 
limit and to extrapolate to the physical pion mass and (c) to take the 
infinite-volume limit. The latter is probed by the infrared behavior of 
gluon and ghost propagators, and it turns out that the approach is extremely 
slow. Discretization effects also show up in the data, but can be easily 
tamed by suitable momentum cuts.
The effects of the vortex mechanism of quark confinement and of dynamical 
quarks on the form of the propagators are very interesting and worth to be
microscopically understood.

\section*{Acknowledgements}
To the organizers of the School-Seminar, in particular to Vladimir Skalozub, I am very 
grateful for inviting me and sponsoring my participation. I enjoyed the inspiring 
meeting and the interesting place. I wish to thank my co-authors A. Sternbeck, 
M. M\"uller-Preussker, I. Bogolubsky, V. Bornyakov, G. Burgio, B. Martemyanov, 
V. Mitrjushkin, A. Schiller and A. Voigt for discussions on the ideas reported here. 
Thanks go to the HLRN regional computing center Berlin/Hannover and to
the Joint Supercomputer Center (JSCC) Moscow for the computing resources.
The support by DFG for my work, presently via the Forschergruppe FOR465, is 
highly appreciated.



\begin{thebibliography}{99}

\bibitem{Alkofer:2006fu}
R. Alkofer and J. Greensite, J. Phys. {\bf G 34}, S3 (2007), arXiv:hep-ph/0610365.
\bibitem{Gribov:1977wm} 
V. N. Gribov, Nucl. Phys. {\bf B 139}, 1 (1978).
\bibitem{Zwanziger:1991ac}
D. Zwanziger, Nucl. Phys. {\bf B 378}, 525 (1992), Nucl. Phys. {\bf B 364}, 127 (1991), Nucl. Phys. {\bf B 412}, 657 (1994).
\bibitem{Kugo:1979gm}
T. Kugo and I. Ojima, Prog. Theor. Phys. Suppl. {\bf 66}, 1 (1979), T. Kugo, (1995), arXiv:hep-th/9511033. 
\bibitem{Alkofer:2000wg}
R. Alkofer and L. von Smekal, Phys. Rept. {\bf 353}, 281 (2001), arXiv:hep-ph/0007355.
\bibitem{Mandula:1987rh}
J. E. Mandula and M. Ogilvie, Phys. Lett. {\bf 185}, 127 (1987).
\bibitem{Mandula:1999nj}
J. E. Mandula, Phys. Rept. {\bf 315}, 273 (1999), arXiv:hep-lat/9907020.
\bibitem{Smekal:1997is}
L. von Smekal, R. Alkofer and A. Hauck, Phys. Rev. Lett. {\bf 78}, 3591 (1997), arXiv:hep-ph/9705242.
\bibitem{Smekal:1998vx}
L. von Smekal, A. Hauck and R. Alkofer, Ann. Phys. {\bf 267}, 1 (1998), arXiv:hep-ph/9707327.
\bibitem{Suman:1995zg}
H. Suman and K. Schilling, Phys. Lett. {\bf B 373}, 314 (1996), arXiv:hep-lat/9512003.
\bibitem{Fischer:2005ui}
C. S. Fischer, B. Gr\"uter and R. Alkofer, Ann. Phys. {\bf 321}, 1918 (2006), arXiv:hep-ph/0506053.
\bibitem{Fischer:2007pf}
C. S. Fischer et al., arXiv:hep-ph/0701050 (2007), Ann. of Phys., in print.
\bibitem{Lerche:2002ep}
C. Lerche and L. von Smekal, Phys. Rev. {\bf D 65},125006 (2002), arXiv:hep-ph/0202194.
\bibitem{Watson:2001yv}
P. Watson and R. Alkofer, Phys. Rev. Lett. {\bf 86}, 5239 (2001), arXiv:hep-ph/0102332.
\bibitem{Zwanziger:2001kw} 
D. Zwanziger, Phys. Rev. {\bf D 65}, 094039 (2002), arXiv:hep-th/0109224.
\bibitem{Zwanziger:2002ia} 
D. Zwanziger, Phys. Rev. {\bf D 67}, 105001 (2003), arXiv:hep-th/0206053.
\bibitem{Bernard:1992hy}
C. W. Bernard, C. Parrinello and A. Soni, Nucl. Phys. Proc. Suppl. {\bf 30}, 535 (1993), arXiv:hep-lat/9211020.
\bibitem{Marenzoni:1993td}
P. Marenzoni et al., Phys. Lett. {\bf B 318}, 511 (1993).
\bibitem{Zwanziger:2003cf}
D. Zwanziger, Phys. Rev. {\bf D 69}, 016002 (2004), arXiv:hep-ph/0303028.
\bibitem{Bali:1996dm}
G. S. Bali et al., Phys. Rev. {\bf D 54}, 2863 (1996), arXiv:hep-lat/9603012.
\bibitem{Schemel:2007xx}
E.-M. Ilgenfritz, M. M\"uller-Preussker and P. Schemel, to be published (2007).
\bibitem{Taylor:1971ff}
J. C. Taylor, Nucl. Phys. {\bf B 33}, 436 (1971).
\bibitem{Bogolubsky:2007bw}
I. L. Bogolubsky, V. G. Bornyakov, G. Burgio, E.-M. Ilgenfritz, M. M\"uller-Preussker and V. K. Mitrjushkin, arXiv:0707.3611 [hep-lat] (2007).
\bibitem{Bogolubsky:2005wf}
I. L. Bogolubsky et al., Phys. Rev. {\bf D 74}, 034503 (2006), arXiv:hep-lat/0511056.
\bibitem{Sternbeck:2005tk}
A. Sternbeck et al., Phys. Rev. {\bf D 72}, 014507 (2005), arXiv:hep-lat/0506007.
\bibitem{Sternbeck:2005re}
A. Sternbeck et al., Nucl. Phys. Proc. Suppl. {\bf 153}, 185 (2006), arXiv:hep-lat/0511053.
\bibitem{Ilgenfritz:2006gp}
E.-M. Ilgenfritz et al., in {\it Sense of Beauty in Physics}, ed. by M. D'Elia et al., Pisa 2006, p. 359, arXiv:hep-lat/0601027 (2006).
\bibitem{Bowman:2007du}
P. O. Bowman et al., arXiv:hep-lat/0703022 (2007).
\bibitem{Bali:2005bg}
G. S. Bali et al., Nucl. Phys. Proc. Suppl. {\bf 153}, 9 (2006) arXiv:hep-lat/0512018.
\bibitem{Boyko:2006ic}
P. Y. Boyko et al., Nucl. Phys. {\bf B 756}, 71 (2006), arXiv:hep-lat/0607003.
\bibitem{Bornyakov:2007fz}
V. G. Bornyakov et al., arXiv:0708.3335 [hep-lat] (2007).
\bibitem{Bornyakov:2003vx}
V. G. Bornyakov et al., Phys. Rev. {\bf D 70}, 074511 (2004), arXiv:hep-lat/0310011.
\bibitem{Langfeld:2001cz}
K. Langfeld, H. Reinhardt and J. Gattnar, Nucl. Phys. {\bf B 621}, 131 (2002), arXiv:hep-ph/0107141.
\bibitem{Gattnar:2004bf}
J. Gattnar, K. Langfeld and H. Reinhardt, Phys. Rev. Lett. {\bf 93}, 061601 (2004), arXiv:hep-lat/0403011.	 
\bibitem{deForcrand:1999ms}
P. de Forcrand and M. D'Elia, Phys. Rev. Lett. {\bf 82}, 4582 (1999), arXiv:hep-lat/9901020.
\bibitem{Greensite:2004ur}
J. Greensite, S. Olejnik and D. Zwanziger, JHEP {\bf 05}, 070 (2005), arXiv:hep-lat/0407032.
\bibitem{Maas:2006ss}
A. Maas, Nucl. Phys. {\bf A 790}, 566 (2007), arXiv:hep-th/0610011.      
\bibitem{Sternbeck:2005vs}
A. Sternbeck, E.-M. Ilgenfritz and M. M\"uller-Preussker, Phys. Rev. {\bf D 73}, 014502 (2006), arXiv:hep-lat/0510109.
\bibitem{Bogolubsky_poster}
I. L. Bogolubsky et al., poster presented at Lattice 2007.
\bibitem{Sternbeck:2006cg}
A. Sternbeck et al., PoS {\bf LAT2006}, 076 (2006), arXiv:hep-lat/0610053 
\bibitem{Ilgenfritz:2006he}
E.-M. Ilgenfritz et al., Braz. J. Phys. {\bf 37}, 193 (2007), arXiv:hep-lat/0609043.
\bibitem{Bloch:2003sk}
J. C. R. Bloch et al., Nucl. Phys. {\bf B 687}, 76 (2004), arXiv:hep-lat/0312036.     

\end{thebibliography}

\end{document}